\documentclass[sigconf,unicode,bookmarks=false]{acmart}

\AtBeginDocument{%
  \providecommand\BibTeX{{%
    \normalfont B\kern-0.5em{\scshape i\kern-0.25em b}\kern-0.8em\TeX}}}

\copyrightyear{2020}
\acmYear{2020}
\setcopyright{rightsretained}
\acmConference[ICSEW'20]{IEEE/ACM 42nd International Conference on Software Engineering Workshops }{May 23--29, 2020}{Seoul, Republic of Korea}
\acmBooktitle{IEEE/ACM 42nd International Conference on Software Engineering Workshops (ICSEW'20), May 23--29, 2020, Seoul, Republic of Korea}
\acmDOI{10.1145/3387940.3392168}
\acmISBN{978-1-4503-7963-2/20/05}

\usepackage{xcolor}
\usepackage[inline]{enumitem}
\usepackage{verbatim}
\usepackage{booktabs} 
\usepackage[capitalise,noabbrev]{cleveref} 
\usepackage{tabularx}
\usepackage{xcolor,colortbl}
\usepackage{comment}
\usepackage{enumitem}

\begin{document}

\title{Mining for Process Improvements:\\Analyzing Software Repositories in Agile Retrospectives}

\author{Christoph Matthies}
\email{christoph.matthies@hpi.de}
\orcid{0000-0002-6612-5055}
\affiliation{%
  \institution{Hasso Plattner Institute}
  \city{University of Potsdam, Germany}
}

\author{Franziska Dobrigkeit}
\email{franziska.dobrigkeit@hpi.de}
\orcid{0000-0001-9039-8777}
\affiliation{%
  \institution{Hasso Plattner Institute}
  \city{University of Potsdam, Germany}
}

\author{Guenter Hesse}
\email{guenter.hesse@hpi.de}
\orcid{0000-0002-7634-3021}
\affiliation{%
  \institution{Hasso Plattner Institute}
  \city{University of Potsdam, Germany}
}

\renewcommand{\shortauthors}{Matthies, Dobrigkeit, Hesse}

\begin{abstract}
Software Repositories contain knowledge on how software engineering teams work, communicate, and collaborate.
It can be used to develop a data-informed view of a team's development process, which in turn can be employed for process improvement initiatives.
In modern, Agile development methods, process improvement takes place in Retrospective meetings, in which the last development iteration is discussed.
However, previously proposed activities that take place in these meetings often do not rely on project data, instead depending solely on the perceptions of team members.
We propose new Retrospective activities, based on \emph{mining} the software repositories of individual teams, to complement existing approaches with more objective, data-informed process views.
\end{abstract}

\ccsdesc[500]{Software and its engineering~Agile software development}

\keywords{Agile Software Development, Software Process Improvement, Mining Software Repositories, Retrospective}

\maketitle

\section{Introduction}
Retrospective meetings are commonly held at the end of a project to review the past work and to identify improvement opportunities.
The practice of Retrospectives was embraced by the Agile community, which focuses on light-weight software development methods, iterations, and feedback~\cite{fowler2001agile}.
Instead of waiting until the end of a project, Agile practitioners began running Retrospective meetings more frequently, e.g. at the end of Scrum Sprints~\cite{Kniberg2015}.
Today, regular Retrospective meetings are a popular practice in professional software engineering~\cite{ScrumAlliance2018}.

\begin{table*}[htb]
    \caption{Extract of types of tools that produce project data which can be employed in data-informed Retrospective activities.}

    \vspace{-0.25em}

    \label{tab:sources}
    \begin{tabularx}{\textwidth}{@{}lXXl@{}}
        \toprule
        \textbf{Tool Type} & \textbf{Function} & \textbf{Examples of Extractable Data Points} & \textbf{Tool Example}\\
        \midrule
        Version Control & Track code changes, communicate rationales~\cite{Santos2016} & Code diffs, committer details, timestamps & \emph{git} \\
        Issue Tracker & Manage detailed information on work items~\cite{Ortu2015} & Developer assignments, status updates & \emph{Jira} \\
        Software Tests & Present the status of current software builds~\cite{Beller2017} & Integration logs, test run logs, build status & \emph{Jenkins} \\
        Status Monitor & Inform/alert regarding availability of systems~\cite{haberkorn2007} & Accumulated uptime, downtime events & \emph{Nagios} \\
        Code Review & Share knowledge, gather critique of peers~\cite{Yang2016} & Time to completion, reviewer details, verdicts & \emph{Gerrit} \\
        Code Analysis & Provide automated feedback on code quality~\cite{williams2005} & Code coverage results, coding style checks & \emph{Lint} \\
        \bottomrule
    \end{tabularx}
\end{table*}

\section{Data-Informed Retro Activities}
Team activities for Retrospectives have been proposed to structure meetings and to encourage the sharing of ideas~\cite{Derby2006}.
Derby and Larsen defined five consecutive phases for Retrospectives in software engineering: \emph{set the stage}, \emph{gather data}, \emph{generate insights}, \emph{decide what to do}, and \emph{close}~\cite{Derby2006}.
More recently, Baldauf introduced the \emph{Retromat}, a book~\cite{Baldauf2018} and online tool, which includes most of the previously proposed exercises in a structured format.
Most proposed Retrospective exercises focus on gathering the perceptions and experiences of team members and extracting improvement opportunities from them.
Another view of the project reality is available through the artifacts that are produced by software developers in the course of their daily work~\cite{Matthies2018b}.
\Cref{tab:sources} lists an extract of popular tools and the data that can be extracted from them.
This data is useful for process improvement as it provides evidence for project problems, e.g. when tests fail~\cite{Ziftci17}.
Large-scale analysis of this valuable project data is the focus of the \emph{Mining Software Repositories} (MSR) research field~\cite{Hassan2008}.
However, their approaches to extract insights from vast collections of software repositories have not yet been applied to software process improvement in small, Agile teams.
We propose employing the software project data of development teams, to enable an additional, data-informed view of the executed process in Retrospective meetings.
Our vision includes new activities for the \emph{gather data} phase, based on software repository analyses.

In the following, we present two use cases: (i) \emph{Action Item Discovery}, i.e. discovering opportunities for improvement and (ii) \emph{Progress Check}, i.e. assessing the team's progress on improvement actions.

\vspace{-0.4em}

\subsection{Action Item Discovery}
The outcome of a Retrospective is a list of ``action items''~\cite{Derby2006}, that the team will work on in the next development iteration.
Of the many proposed activities to gather data, only extremely few have a connection to project data~\cite{Baldauf2018}.
We propose using data-driven activities to discover new action items.
Assessments of project data can be drawn from measurements designed for Agile software engineering best practices.
Examples include code coverage over time,~\cite{Estacio2014}, the regularity of commits to the VCS~\cite{Matthies2016} or the percentage of stories implemented using Pair Programming~\cite{Estacio2014}.

\vspace{-0.2em}

\subsubsection*{Proposed Activity: Health Check}
The Retrospective exercise is based on the established software development best practices of a team's organization, with the goal of revealing violations of these practices in the project data.
To \emph{gather data}, project data measurements concerning a practice should be collected.
For example, for the ``commit early, commit often'' principle~\cite{atwood08}, this can include the average amount of commits per developer or the average time between commits during core working hours.
In the \emph{generate insights} phase, the team members can inspect the results and note whether they are outside the expected range, i.e. when adhering to the rule.
The team members can compare their interpretations of analysis results, debate rationales for their observations and can find a consensus on action items for the next iteration, e.g. to commit their work to the VCS after each finished work item.
In the case that results are considered to be flawed or false positives, the measurement parameters can be fine-tuned for the next iteration.

\subsection{Progress Check}
Without a method to gain insight into the effectiveness of Retrospectives and few tangible results, an organization might find it hard to justify the time and expense of performing Retrospectives~\cite{Marshburn2018}.
Project artifact measurements, based on Retrospective action items, are one avenue to provide these quantifiable improvement results.
Once a measurement is defined for a given action item, the results for the current (without the change) and the next iteration (with the enacted change) can be compared.

\subsubsection*{Proposed Activity: Remedy Appraisal}
Suppose that in a previous Retrospective the team identified the issue of a single person committing most of the team's code changes, which slowed down the team.
As an action item, all team members were trained in VCS usage.
To track progress, the team can decide to employ the number of unique contributors to their code repository as a measurement.
In the following Retrospective, the team appraises the effect of the remedy.
The VCS can provide evidence of whether the training showed effects and whether more team members contributed code, by rerunning the previously defined measurements and comparing results.
Depending on whether the results improve, i.e. show a higher contributor count, the action item can be considered resolved or can be discussed further.

\section{Conclusion}
Modern software engineers depend on digital collaboration, communication and development tools.
Integrations between these tools are becoming more prevalent.
An increasing amount of information on developers' interactions and behaviors is available in project artifacts, which allows improving cooperative and development processes~\cite{Singer2017}.
However, these concepts have not yet fully established themselves in the domain of Agile process improvement.
We propose new Retrospective activities based on project data measurements both for discovering process improvement opportunities and progress inspection.
Our proposal represents initial steps in integrating the promises of the field of \emph{Mining Software Repositories} into Agile process improvement approaches.
Future work includes research on automating data-informed insights, such as through chatbots supporting Agile Retrospectives~\cite{Matthies2019BotSE}.

\bibliographystyle{ACM-Reference-Format}
\bibliography{library}

\end{document}